\documentclass{optica-article}

\journal{opticajournal} 

\articletype{Research Article}

\usepackage{lineno}

\usepackage[separate-uncertainty=true]{siunitx}
\usepackage{xcolor}
\usepackage{gensymb}

\begin{document}

\title{Spatial mapping and tuning of terahertz modes in a silicon whispering gallery mode resonator}

\author{Anna R. Petersen\authormark{1,2}, Pablo Paulsen\authormark{1,2}, Florian Sedlmeir\authormark{1,2}, Nicholas J. Lambert\authormark{1,2}, Harald G. L. Schwefel\authormark{1,2}, and Mallika Irene Suresh\authormark{1,2,*}}

\address{\authormark{1}Department of Physics, University of Otago, Dunedin, New Zealand\\
\authormark{2}Dodd-Walls Centre for Photonic and Quantum Technologies, New Zealand\\
}
\email{\authormark{*}mallika.suresh@otago.ac.nz} 

\begin{abstract*}  
Identification and subsequent manipulation of resonant modes typically relies on comparison with calculations which require precise knowledge of material parameters and dimensions. For millimetre-sized resonators that support millimetre-wave modes, this is particularly challenging due to the poorly determined physical parameters and additional perturbative effects of nearby dielectric or metallic substrates in experimental setups. Here, we use perturbation from a metal needle to experimentally map the spatial distribution of terahertz modes in a silicon disc resonator. We then use this information to design a patterned structure to manipulate specific modes in the system, a technique that could be useful for targeted tuning of such modes.
\end{abstract*}

\section{Introduction}
Dielectric resonators are structures in which resonant electromagnetic modes are supported due to confinement by reflection at the dielectric-air interface. They have found use at radiofrequency and microwave frequencies~\cite{huitema_dielectric_2012,noori_microwave_2018,dey_low-loss_2021,bark_transmission_2018}, with many applications leveraging the lower loss they exhibit compared to their metal equivalents~\cite{devi_performance_2019,alanazi_review_2023}. At optical frequencies, repeated total internal reflection at the boundary of a disc or sphere leads to the formation of whispering gallery modes (WGMs), which can exhibit high quality factors. WGMs have a variety of applications~\cite{matsko_review_2005,lin_nonlinear_2017}, such as sensing~\cite{toropov_review_2021,yu_whispering-gallery-mode_2021,loyez_whispering_2023}, and, in materials exhibiting second and third order nonlinearities, frequency conversion via either three- or four-wave mixing~\cite{matsko_highly_2002,strekalov_efficient_2009,strekalov_nonlinear_2016,breunig_three-wave_2016,liu_nonlinear_2021,lambert_coherent_2020, lambert_microresonator-based_2023}. WGM resonators also have applications at THz frequencies, and have been demonstrated as tunable filters~\cite{vogt_thermal_2018, wang_voltage-actuated_2019, xie_terahertz-frequency_2020, gupta_electrically_2023}, sensors~\cite{mathai_sensing_2018,vogt_terahertz_2020,hou_crystalline_2022}, isolators~\cite{yuan_-chip_2021} and frequency references~\cite{gandhi_microresonator_2022}. 

To use resonators most effectively, it is often important to have an accurate knowledge of the spatial distribution of the modes supported therein. This is particularly crucial in experiments that require careful design of the geometry of such cavities to achieve phase-matching for nonlinear optical processes~\cite{ilchenko_whispering-gallery-mode_2003,soltani_efficient_2017,botello_sensitivity_2018, abdalmalak_integrated_2022,logan_triply-resonant_2023,suresh_multichannel_2025}. In the case of WGMs, the spatial distribution is characterized by three mode numbers: the radial number $q$ (the number of maxima in the radial direction); the polar number $p$ (the number of nodes in the polar direction) and the azimuthal number $m$ (the number of wavelengths that fit in the circumference of the resonator). In simple spherical geometries, closed form solutions exist for the spatial distribution and frequencies of the resonant fields. However, in more complex geometries these usually must be determined by finite element methods.

While these simulations often provide a good estimate of each mode's central frequency and spatial distribution, their accuracy is limited by uncertainties in geometry and material parameters, and the effects of external perturbations which may be difficult to take into account. This can cause large discrepancies between the numerical solutions and experiment. This motivates the development of methods to experimentally determine each mode's spatial form. Work in this direction has been done by measuring the far-field of optical WGMs~\cite{schunk_identifying_2014} as well as near-field microscopy~\cite{knight_mapping_1995,schmidt_near-field_2012}. Near-field imaging has also been used more recently to map microwave modes~\cite{dev_near-field_2022} and THz modes~\cite{hale_near-field_2023} in periodic dielectric resonators~\cite{lee_terahertz_2017}, plasmonic resonators~\cite{mitrofanov_near-field_2018}, and metallic ring resonators~\cite{schiattarella_terahertz_2024}. 

A notable feature of dielectric resonators is that, unlike metallic cavity resonators, a portion of the modes' field extends outside the resonator. This means that the mode frequency can be tuned or the mode form manipulated by bringing external structures (typically dielectric or metallic) into the field~\cite{foreman_dielectric_2016, azeem_dielectric_2021,vogt_anomalous_2019, suresh_gallium_2023}, changing the boundary conditions of the mode. Such strategies are a useful tool, especially in frequency conversion experiments where optimized phase-matching of interacting signals is required~\cite{ta_tuning_2013,schunk_frequency_2016,munoz-hernandez_tunable_2019,abdalmalak_integrated_2022}.

In this paper, we demonstrate the use of this perturbative effect in a system in which external perturbations of the field by nearby dielectrics is particularly noticeable. We consider the WGMs of a millimetre-sized disc resonator in the D-band (\qtyrange{130}{174.8}{GHz}). In this regime, the resonator dimensions are of similar size to the wavelength of the THz radiation forming the WGMs, resulting in a large external field. We use the localised perturbative effect of a stainless steel needle tip scanned across the resonator to map the spatial distribution of THz frequency modes. Using this information, we then design and fabricate a structure to perturb the external field, selectively tuning and frequency splitting a specific THz mode.

\section{Experiment}
\begin{figure}[htbp]
\centering\includegraphics[width = 13cm]{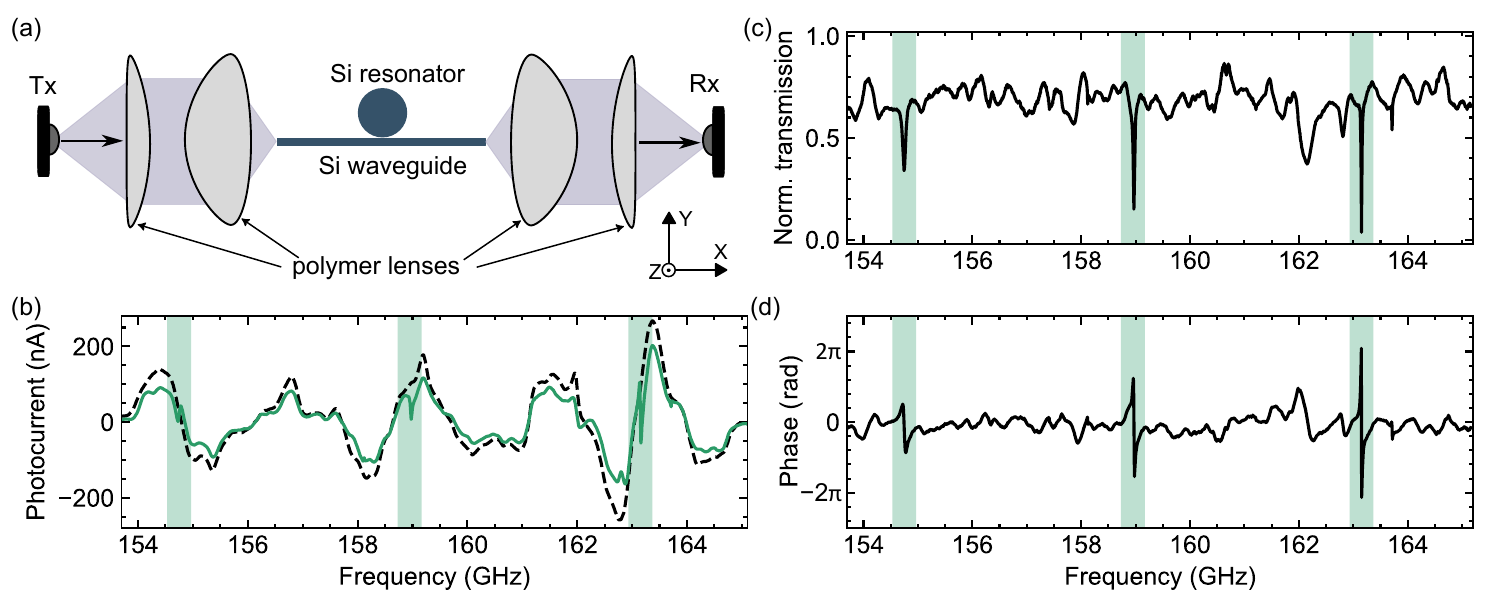}
\caption{THz spectroscopy system: (a) Schematic showing two pairs of polymer lenses focussing the THz radiation from transmitter (Tx) and to receiver (Rx) into a silicon (Si) waveguide coupled to a Si disc resonator. (b) Measured photocurrent at Rx showing a reference spectrum with no resonator (black dashed curve) and the spectrum with the resonator present (green curve). From the photocurrent we extract the (c) amplitude and (d) phase components of the spectrum via a Hilbert transformation. The three observed modes are highlighted in green.}
\label{fig:simplesetup}
\end{figure}

We use a silicon disc resonator with a radius of \SI{2.89}{mm} and thickness of \SI{354}{\micro m}. To excite WGMs, tunable continuous-wave (CW) THz radiation is generated by a commercially-available TeraScan system (Toptica Photonics)~\cite{stanze_compact_2011, deninger_275_2015}. The radiation is collimated and focused down to a silicon waveguide (with rectangular cross-section: height \SI{0.35}{mm}, width \SI{0.2}{mm}) by polymer lenses. It is then coupled into the resonator via frustrated internal reflection (Fig.~\ref{fig:simplesetup}(a)). The polarisation of the radiation is chosen such that we excite vertically-polarised WGMs (along the Z-axis in Fig.~\ref{fig:simplesetup}(a)). 

We use a coherent detection scheme, measuring the photocurrent created by mixing the THz signal at the receiver (Rx) with the optical beat signal used for the THz generation at the transmitter (Tx). The photocurrent is therefore proportional to the amplitude of the electric field of the THz radiation modulated by the phase difference between the THz radiation and the optical beat signal. We measure the absorption spectrum of the Si resonator as follows: we acquire reference and sample spectra with and without the disc resonator coupled to the waveguide (black dashed curve and green curve respectively in Fig.~\ref{fig:simplesetup}(b)). Using a Hilbert transformation~\cite{vogt_high_2017}, we extract the amplitude and phase components of the transmission spectrum (Figs \ref{fig:simplesetup}(c) and (d) respectively). We observe resonances (highlighted in green) at \SI{154.75}{GHz}, \SI{158.97}{GHz} and \SI{163.15}{GHz}.

\subsection{Mapping of the spatial distribution of modes}
\label{subsection:mapping}
We now aim to image the spatial profile of the modes in our system. While the eigenmodes of a perfectly rotationally-symmetric WGM resonator are travelling waves, random inhomogeneities of the bulk or surface~\cite{mazzei_controlled_2007}, or (more significantly in the THz case), the presence of nearby dielectric components such as a coupling structure can break the rotational symmetry and couple the clockwise and counter-clockwise propagating waves via back scattering. This results in the eigenmodes being a pair of standing wave modes. 

\begin{figure}[htbp]
\centering\includegraphics[width = 13cm]{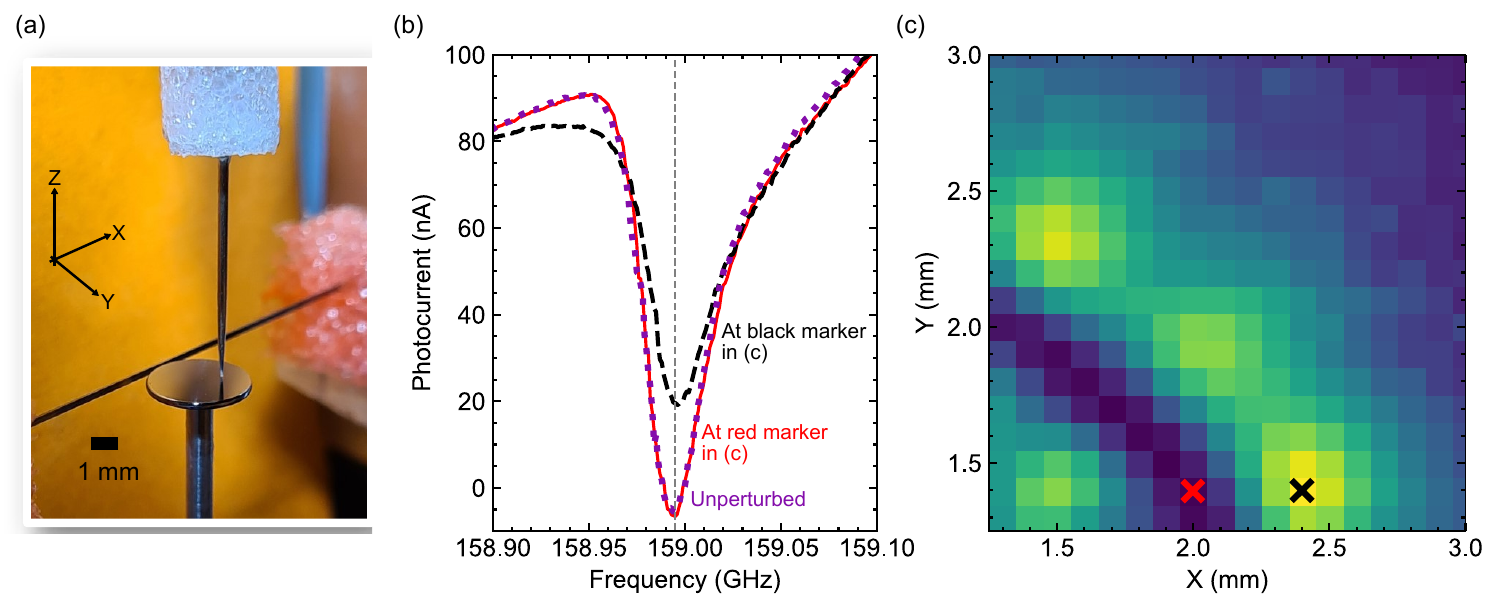}
\caption{Working principle: (a) Photograph of the needle probe mounted above the resonator and waveguide. (b) Photocurrent spectra when the needle is far from the resonator (purple dotted curve), at a node of the electric field (red continuous curve) and at an antinode of the electric field, leading to perturbation of the mode by the needle (black dashed curve). (c) A map generated by scanning the needle over a part of the resonator, with the frequency fixed at \SI{158.995}{GHz} (vertical grey line in (b)), and plotting the resulting photocurrent.}
\label{fig:mappingmethods}
\end{figure}

We map the field structure of the modes we excite by introducing the tip of a stainless steel needle into the portion of the mode residing in air above the silicon resonator. The \qty{0.1(0.05)}{\milli\metre} diameter needle tip (mounted in foam and fixed to a 3D translation stage) is oriented perpendicular to the top of the resonator (Fig.~\ref{fig:mappingmethods}(a)) at a distance of \qty{0.1(0.05)}{\milli\metre}. If the tip is at a maximum of the standing wave field, the loss rate of the mode increases in comparison to the case where the tip is at a minimum of the field, in a similar way to the method used for optical modes demonstrated in~\cite{wang_direct_2021}. The increase in loss rate changes the coupling contrast of the mode (Fig.~\ref{fig:mappingmethods}(b)). We detect the degree of perturbation by fixing the THz frequency at that of the unperturbed mode and measuring the change in photocurrent with the needle at different positions. As the photocurrent is proportional to the electric field amplitude of the THz field being  measured at the receiver, we can deduce the electric field distribution of the excited mode (Fig.~\ref{fig:mappingmethods}(c)).

\begin{figure}[htbp!]
\centering\includegraphics[width=13cm]{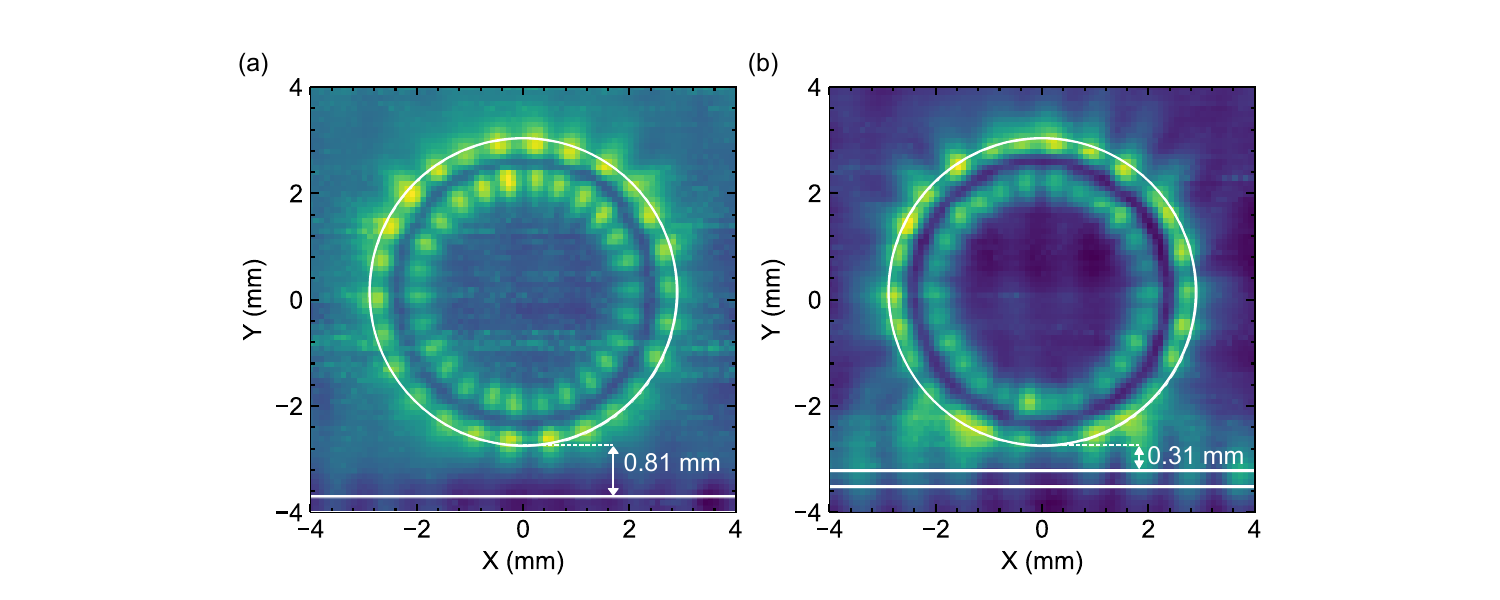}
\caption{Maps of the mode centred at \SI{158.97}{GHz} with $m=13$, at different distances between the resonator and the waveguide: (a) \SI{0.81}{mm} and (b) \SI{0.31}{mm}.}
\label{fig:couplingmaps}
\end{figure}

First, we examine the structure of the mode centred at \qty{158.97}{\giga\hertz}. We proceed by exciting the mode in the resonator, and scanning the probe needle across the plane above the resonator and waveguide. In Fig.~\ref{fig:couplingmaps}, we show the measured photocurrent when the waveguide is \qty{0.81}{\milli\metre} away from the resonator (panel (a)), and when the waveguide is \qty{0.31}{\milli\metre} from the resonator (panel (b)). In Fig.~\ref{fig:couplingmaps}(a), the spatial distribution of the field exhibits a high degree of rotational symmetry, indicating that the waveguide has a negligible effect on the mode structure. In contrast, in Fig.~\ref{fig:couplingmaps}(b) the coupling dielectric structure has a significant effect on the shape of the mode. In both cases, the mode structure is clearly visible; we identify it to have indices $q=2$ and $m=13$, due to the 2 peaks in the radial and 26 peaks in the azimuthal direction. We also note that the standing wave that we excite has a minimum at the point nearest to the waveguide regardless of coupling strength, due to the nature of the coupling interaction. 

\begin{figure}[htbp]
\centering\includegraphics[width = 14cm]{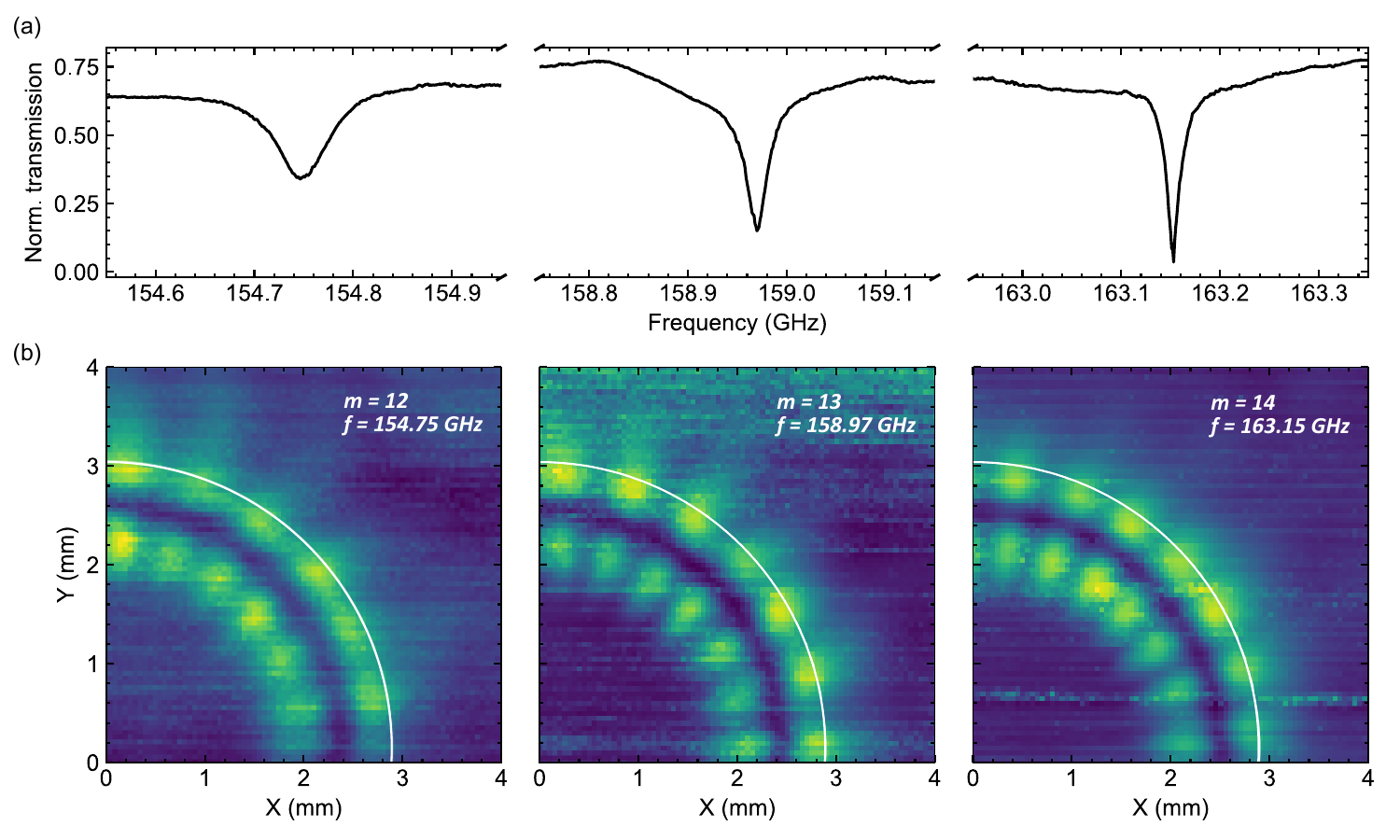}
\caption{Identifying three consecutive azimuthal order modes: (a) Normalised transmission spectra and (b) quarter maps of the three modes highlighted in Fig.~\ref{fig:simplesetup}: (left) $m=12$, $f=\SI{154.75}{GHz}$, (middle) $m=13$, $f=\SI{158.97}{GHz}$, (right) $m=14$, $f=\SI{163.15}{GHz}$.}
\label{fig:diffazimuthal}
\end{figure}

In a similar manner, we also image the two other modes highlighted in green in Fig.~\ref{fig:simplesetup}. We show the quarter maps of all three modes in Fig.~\ref{fig:diffazimuthal}(b), and identify these resonances as the consecutive $q=2$ modes, with the mode centred on $f=\SI{154.75}{GHz}$ having $m=12$ and that at $f=\SI{163.15}{GHz}$ having $m=14$. To allow for higher resolution maps to be taken within the same time, here we measure quarter maps with each map taking approximately 4 hours (limited by the settling time of the stages).

\subsection{Manipulation of targeted modes}
\label{subsection:splitting}
\begin{figure}[htbp]
\centering\includegraphics[width = 13cm]{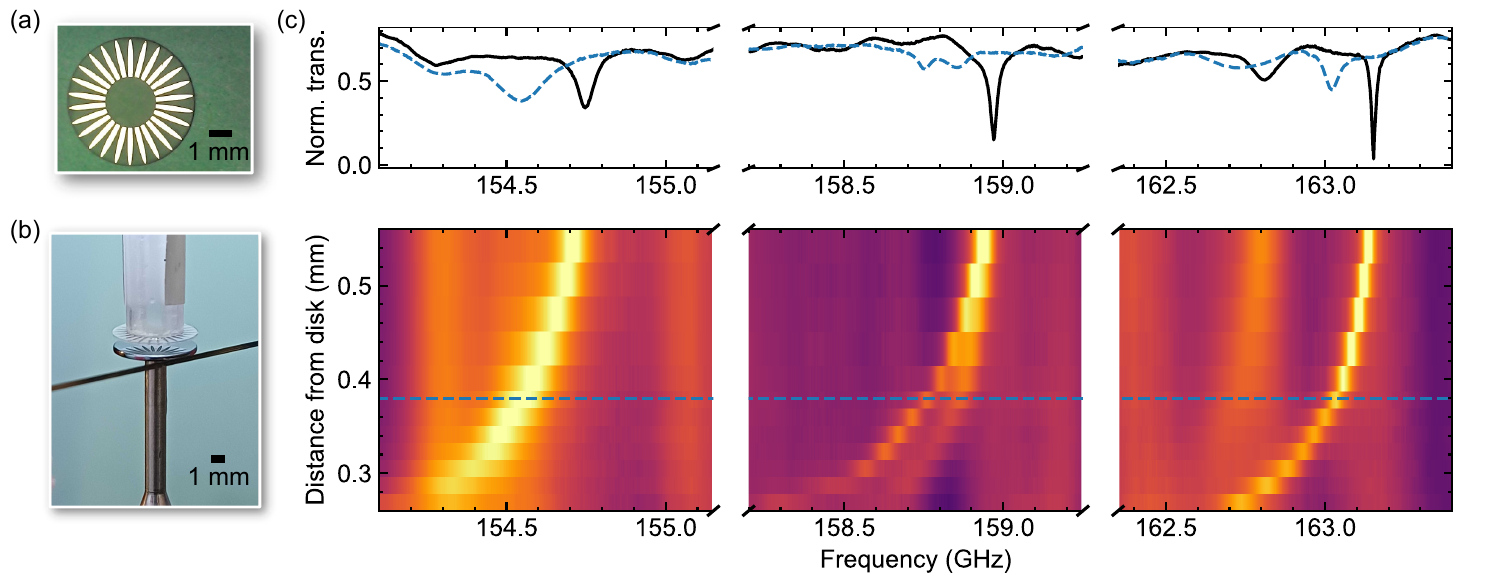}
\caption{Targeted mode manipulation: (a) Photograph of the perturbative structure showing the 26 metal lobes on a thin glass plate. (b) Photograph showing the perturbative structure above the Si resonator coupled to the waveguide. (c) (above) Normalized transmission spectra of the unperturbed modes (plate $\gg\SI{0.6}{mm}$ above the resonator, black curve) and perturbed modes (plate \SI{0.38}{mm} above the resonator, blue dashed curves). (below) Mode spectra as a function of distance of the patterned plate from the resonator, showing the red-shift of all three modes, and the splitting of the $m=13$ mode. Blue dashed lines correspond to the blue dashed curves in the upper panel.}
\label{fig:distancesweep}
\end{figure}

In this section, we use the information gained from the electric field maps described above for targeted manipulation of a specific mode. We rely on the symmetry breaking induced by a periodic change in permittivity (similar to that present in photonic crystals), which lifts the degeneracy between standing waves with a commensurate spatial period and opens an energy gap between them. We again exploit the significant part of the field outside the silicon resonator by perturbing the mode from above with a rotationally-periodic structure (Fig.~\ref{fig:distancesweep}(a)).

The perturbative structure is fabricated by coating a microscope cover slip (made of fused silica glass) with \SI{50}{nm} of chromium followed by \SI{120}{nm} of silver in a thermal evaporator. The metal is then patterned using femtosecond laser ablation, and the resulting structure cut out from the cover slip such that it is centred on a circular glass plate of radius \SI{2.9}{mm}. The periodic lobed pattern is shown in Fig.~\ref{fig:distancesweep}(a). It has 26 metal lobes equally spaced in the azimuthal direction, and extending radially from $r=\SI{0.5}{mm}$ to the rim of the glass plate; it therefore exhibits rotational symmetry around its axis with period $\ang{360}/26=\ang{13.85}$, matching the spatial periodicity of the $m=13$ mode.

The patterned glass plate is then glued to a narrow plastic cylinder, mounted on a rotation-translation stage and centered above the resonator. The metal pattern is oriented such that the metal lobe closest to the waveguide is aligned with the Y axis. To observe the effect of the perturbing plate on the mode frequencies, the spectrum of the three modes is measured as a function of the distance between the plate and the resonator. The results are shown in Fig.~\ref{fig:distancesweep}(c), with the top panels showing the unperturbed (black curves) and perturbed (dashed blue curves) spectra. The perturbed spectrum is taken with a plate--resonator separation of \SI{0.38}{mm} (marked as horizontal lines in the lower panel).

We observe all three modes shift to lower frequencies and increase in linewidth, suggesting an increase in both effective index and loss rate of the modes. This is consistent with the behaviour seen in simple axisymmetric FEM simulations, which show that a metal or dielectric perturbation above the resonator forces the field to lie in the resonator or in the perturbative dielectric respectively, rather than in air.

\begin{figure}[htbp]
\centering\includegraphics[width = 13cm]{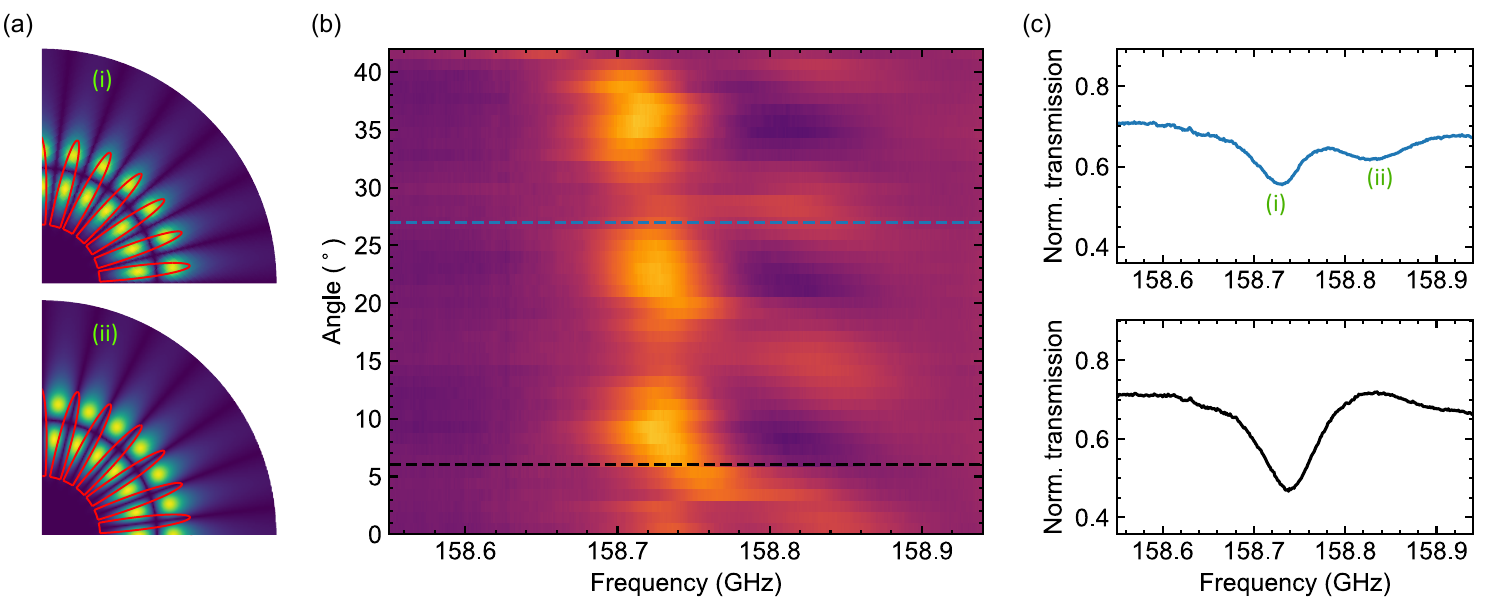}
\caption{Angular dependence of mode perturbation: (a) Two non-degenerate mode patterns as calculated in a simple axisymmetric FEM simulation in COMSOL Multiphysics\textsuperscript\textregistered showing the positions of the field maxima relative to the lobe pattern (outlined in red) (i) standing wave maxima aligned with the metal lobes and (ii) standing wave maxima between the metal lobes. (b) Transmission spectra as a function of angle between lobes and standing wave. (c) Spectra at positions of horizontal lines in (b) showing single and split modes.}
\label{fig:anglesweep}
\end{figure}

We also find that the $m=13$ mode ($f\approx\SI{158.97}{GHz}$), which has a periodicity commensurate with the patterned plate, splits into two distinct resonances. We ascribe this to the reduction in rotational symmetry leading to the lifting of the degeneracy between the mode having standing wave maxima aligned with the metal lobes (Fig.~\ref{fig:anglesweep}(a), panel (i)), and the mode having its maxima aligned with the unscreened silica sections in between the metal lobes (Fig.~\ref{fig:anglesweep}(a), panel (ii)). From axisymmetric FEM simulations, we know that a metal perturbation from above forces the field of the mode to reside in the silicon resonator, while a lower-index dielectric (in our case silica) above the resonator causes the field to instead extend into the silica. These two modes therefore have a significant difference in effective refractive index, and hence frequency, resulting in the splitting of the absorption line as the patterned plate is brought into proximity with the resonator. We conclude that the mode experiencing the larger redshift has its maxima directly underneath the metal lobes while the higher frequency mode has its maxima aligned with the silica dielectric in between the metal lobes.

To better understand the splitting of the $m=13$ mode, we measure the transmission spectrum as a function of the rotation angle of the metal lobes. Starting with the the metal lobe closest to the waveguide aligned with the Y axis, as for Fig.~\ref{fig:distancesweep}, we rotate the plate over \ang{42} (approximately 3 periods of the metallic pattern) in \ang{1} increments, recording a spectrum for each step. The resulting spectra are displayed in Fig.~\ref{fig:anglesweep}(b).
We observe a spectral dependence on angle with period \ang{13.85}. 
 
At an angle of \ang{0}, each metal lobe is directly above a standing wave minimum. In this orientation, two spectral dips are observed (example marked by upper dashed blue line). As the angle is increased, the higher frequency dip decreases in contrast until it vanishes at \ang{6.9} when each lobe is above a standing wave maximum (example marked by lower dashed black line). The spectra corresponding to the two dashed lines are shown in Fig.~\ref{fig:anglesweep}(c).

These results show that the metal structure can have a significant effect on the coupling mechanism as well as the mode frequencies. When a metal lobe is closest to the coupling point, the wave in the waveguide is perturbed such that it has an overlap with both of the standing wave modes allowing some coupling to both of them ((i) and (ii)) but with reduced contrast. In the case for which the metal lobe is not at this point, the coupling is more like that in Section~\ref{subsection:mapping}, in which we excite only one mode.

\section{Conclusion}
We mapped the field structure of D-band WGMs excited in a silicon disc resonator by scanning a simple metal needle across the mode and measuring the resulting changes in the transmission. This scheme enabled precise experimental mode identification, as well as a method of visualising the effect nearby structures have on the mode form. We expect that this will be a useful tool for the careful dispersion engineering of schemes that require accurate knowledge of the spatial form of the modes, e.g., for phase-matching in nonlinear optical mixing processes~\cite{ilchenko_whispering-gallery-mode_2003,soltani_efficient_2017,botello_sensitivity_2018,abdalmalak_integrated_2022,logan_triply-resonant_2023,suresh_multichannel_2025}. Once the modes are identified, selective manipulation of specific modes would offer an additional degree of freedom for tuning towards optimal phase-matching for such schemes to alleviate constraints on the fabrication of the cavities~\cite{abdalmalak_integrated_2022}. To this end, we explored the use of a perturbative element designed to target and selectively split one of the modes. The structure was patterned to have a periodic structure that matches the symmetry of the targeted mode. Such splitting could also be used as an quicker alternative to identify the azimuthal order of symmetry of modes instead of the comparatively time-consuming process of mapping out the electric field in a large area. We demonstrated that the splitting could be varied as a function of the plate-resonator separation. Furthermore, the coupling of the selected mode can be varied as a function of the angle of the plate relative to the resonator and in-coupling waveguide. 
Our approach is therefore able to tune coupling, frequency and linewidth of modes in a dielectric resonator. The technique is versatile and would allow different perturbative designs to achieve different effects.

\begin{backmatter}
\bmsection{Funding} Dodd-Walls Centre; Ministry of Business, Innovation and Employment.

\bmsection{Disclosures} The authors declare no conflicts of interest.

\bmsection{Data availability} Data underlying the results presented in this paper are available at Mapping terahertz field distributions, Anna R. Petersen, Pablo Paulsen, Florian Sedlmeir, Nicholas J. Lambert, Harald G. L. Schwefel, Mallika Irene Suresh, Zenodo 2025 (10.5281/zenodo.15373139).

\end{backmatter}



\begin{thebibliography}{10}
\newcommand{\enquote}[1]{``#1''}

\bibitem{huitema_dielectric_2012}
L.~Huitema, T.~Monediere, L.~Huitema, and T.~Monediere, \enquote{Dielectric {Materials} for {Compact} {Dielectric} {Resonator} {Antenna} {Applications},} in \emph{Dielectric {Material},}  (IntechOpen, 2012).

\bibitem{noori_microwave_2018}
A.~S. Noori, X.~Shang, C.~Guo, \emph{et~al.}, \enquote{Microwave filters based on novel dielectric split-ring resonators with high unloaded quality factors,} {\protect\JournalTitle{IET Microwaves, Antennas \& Propagation}} \textbf{12}, 1389--1394 (2018).

\bibitem{dey_low-loss_2021}
U.~Dey, J.~G. Marin, and J.~Hesselbarth, \enquote{Low-loss and tunable millimeter-wave filters using spherical dielectric resonators,} {\protect\JournalTitle{International Journal of Microwave and Wireless Technologies}} \textbf{13}, 751--755 (2021).

\bibitem{bark_transmission_2018}
H.~S. Bark, G.~J. Kim, and T.-I. Jeon, \enquote{Transmission characteristics of all-dielectric guided-mode resonance filter in the {THz} region,} {\protect\JournalTitle{Scientific Reports}} \textbf{8}, 13570 (2018).

\bibitem{devi_performance_2019}
K.~R. Devi, \enquote{Performance comparison of microstrip antenna and dielectric resonator antenna ({DRA}) at {RFID} application,} in \emph{Computer-{Aided} {Developments}: {Electronics} and {Communication},}  (CRC Press, 2019).

\bibitem{alanazi_review_2023}
M.~D. Alanazi, \enquote{A {Review} of {Dielectric} {Resonator} {Antenna} at {mm}-{Wave} {Band},} {\protect\JournalTitle{Eng}} \textbf{4}, 843--856 (2023).

\bibitem{matsko_review_2005}
A.~B. Matsko, A.~A. Savchenkov, D.~Strekalov, \emph{et~al.}, \enquote{Review of {Applications} of {Whispering}-{Gallery} {Mode} {Resonators} in {Photonics} and {Nonlinear} {Optics},} {\protect\JournalTitle{Interplanetary Network Progress Report}} \textbf{42-162}, 1--51 (2005).

\bibitem{lin_nonlinear_2017}
G.~Lin, A.~Coillet, and Y.~K. Chembo, \enquote{Nonlinear photonics with high-{Q} whispering-gallery-mode resonators,} {\protect\JournalTitle{Advances in Optics and Photonics}} \textbf{9}, 828--890 (2017). Publisher: Optica Publishing Group.

\bibitem{toropov_review_2021}
N.~Toropov, G.~Cabello, M.~P. Serrano, \emph{et~al.}, \enquote{Review of biosensing with whispering-gallery mode lasers,} {\protect\JournalTitle{Light: Science \& Applications}} \textbf{10}, 42 (2021).

\bibitem{yu_whispering-gallery-mode_2021}
D.~Yu, M.~Humar, K.~Meserve, \emph{et~al.}, \enquote{Whispering-gallery-mode sensors for biological and physical sensing,} {\protect\JournalTitle{Nature Reviews Methods Primers}} \textbf{1}, 1--22 (2021).

\bibitem{loyez_whispering_2023}
M.~Loyez, M.~Adolphson, J.~Liao, and L.~Yang, \enquote{From {Whispering} {Gallery} {Mode} {Resonators} to {Biochemical} {Sensors},} {\protect\JournalTitle{ACS Sensors}} \textbf{8}, 2440--2470 (2023).

\bibitem{matsko_highly_2002}
A.~B. Matsko, V.~S. Ilchenko, A.~A. Savchenkov, and L.~Maleki, \enquote{Highly nondegenerate all-resonant optical parametric oscillator,} {\protect\JournalTitle{Physical Review A}} \textbf{66}, 043814 (2002). Publisher: American Physical Society.

\bibitem{strekalov_efficient_2009}
D.~V. Strekalov, A.~A. Savchenkov, A.~B. Matsko, and N.~Yu, \enquote{Efficient upconversion of subterahertz radiation in a high-{Q} whispering gallery resonator,} {\protect\JournalTitle{Optics Letters}} \textbf{34}, 713--715 (2009). Publisher: Optica Publishing Group.

\bibitem{strekalov_nonlinear_2016}
D.~V. Strekalov, C.~Marquardt, A.~B. Matsko, \emph{et~al.}, \enquote{Nonlinear and quantum optics with whispering gallery resonators,} {\protect\JournalTitle{Journal of Optics}} \textbf{18}, 123002 (2016).

\bibitem{breunig_three-wave_2016}
I.~Breunig, \enquote{Three-wave mixing in whispering gallery resonators,} {\protect\JournalTitle{Laser \& Photonics Reviews}} \textbf{10}, 569--587 (2016).

\bibitem{liu_nonlinear_2021}
W.~Liu, Y.-L. Chen, S.-J. Tang, \emph{et~al.}, \enquote{Nonlinear {Sensing} with {Whispering}-{Gallery} {Mode} {Microcavities}: {From} {Label}-{Free} {Detection} to {Spectral} {Fingerprinting},} {\protect\JournalTitle{Nano Letters}} \textbf{21}, 1566--1575 (2021).

\bibitem{lambert_coherent_2020}
N.~J. Lambert, A.~Rueda, F.~Sedlmeir, and H.~G.~L. Schwefel, \enquote{Coherent {Conversion} {Between} {Microwave} and {Optical} {Photons}—{An} {Overview} of {Physical} {Implementations},} {\protect\JournalTitle{Advanced Quantum Technologies}} \textbf{3}, 1900077 (2020).

\bibitem{lambert_microresonator-based_2023}
N.~J. Lambert, L.~S. Trainor, and H.~G.~L. Schwefel, \enquote{Microresonator-based electro-optic dual frequency comb,} {\protect\JournalTitle{Communications Physics}} \textbf{6}, 1--8 (2023).

\bibitem{vogt_thermal_2018}
D.~W. Vogt, A.~H. Jones, and R.~Leonhardt, \enquote{Thermal tuning of silicon terahertz whispering-gallery mode resonators,} {\protect\JournalTitle{Applied Physics Letters}} \textbf{113}, 011101 (2018).

\bibitem{wang_voltage-actuated_2019}
Z.~Wang, G.~Dong, S.~Yuan, \emph{et~al.}, \enquote{Voltage-actuated thermally tunable on-chip terahertz filters based on a whispering gallery mode resonator,} {\protect\JournalTitle{Optics Letters}} \textbf{44}, 4670--4673 (2019).

\bibitem{xie_terahertz-frequency_2020}
J.~Xie, X.~Zhu, H.~Zhang, \emph{et~al.}, \enquote{Terahertz-frequency temporal differentiator enabled by a high-{Q} resonator,} {\protect\JournalTitle{Optics Express}} \textbf{28}, 7898--7905 (2020).

\bibitem{gupta_electrically_2023}
M.~Gupta, A.~Kumar, and R.~Singh, \enquote{Electrically {Tunable} {Topological} {Notch} {Filter} for {THz} {Integrated} {Photonics},} {\protect\JournalTitle{Advanced Optical Materials}} \textbf{11}, 2301051 (2023).

\bibitem{mathai_sensing_2018}
C.~Mathai, R.~Jain, V.~G. Achanta, \emph{et~al.}, \enquote{Sensing at terahertz frequency domain using a sapphire whispering gallery mode resonator,} {\protect\JournalTitle{Optics Letters}} \textbf{43}, 5383--5386 (2018).

\bibitem{vogt_terahertz_2020}
D.~W. Vogt, A.~H. Jones, and R.~Leonhardt, \enquote{Terahertz {Gas}-{Phase} {Spectroscopy} {Using} a {Sub}-{Wavelength} {Thick} {Ultrahigh}-{Q} {Microresonator},} {\protect\JournalTitle{Sensors}} \textbf{20}, 3005 (2020).

\bibitem{hou_crystalline_2022}
Z.~Hou, S.~Yuan, W.~Deng, \emph{et~al.}, \enquote{Crystalline {Hydrate} {Dehydration} {Sensing} {Based} on {Integrated} {Terahertz} {Whispering} {Gallery} {Mode} {Resonators},} {\protect\JournalTitle{Sensors}} \textbf{22}, 9116 (2022).

\bibitem{yuan_-chip_2021}
S.~Yuan, L.~Chen, Z.~Wang, \emph{et~al.}, \enquote{On-chip terahertz isolator with ultrahigh isolation ratios,} {\protect\JournalTitle{Nature Communications}} \textbf{12}, 5570 (2021).

\bibitem{gandhi_microresonator_2022}
R.~Gandhi, R.~Leonhardt, and D.~W. Vogt, \enquote{Microresonator {Frequency} {Reference} for {Terahertz} {Precision} {Sensing} and {Metrology},} {\protect\JournalTitle{IEEE Transactions on Terahertz Science and Technology}} \textbf{12}, 70--74 (2022).

\bibitem{ilchenko_whispering-gallery-mode_2003}
V.~S. Ilchenko, A.~A. Savchenkov, A.~B. Matsko, and L.~Maleki, \enquote{Whispering-gallery-mode electro-optic modulator and photonic microwave receiver,} {\protect\JournalTitle{JOSA B}} \textbf{20}, 333--342 (2003).

\bibitem{soltani_efficient_2017}
M.~Soltani, M.~Zhang, C.~Ryan, \emph{et~al.}, \enquote{Efficient quantum microwave-to-optical conversion using electro-optic nanophotonic coupled resonators,} {\protect\JournalTitle{Physical Review A}} \textbf{96}, 043808 (2017).

\bibitem{botello_sensitivity_2018}
G.~S. Botello, F.~Sedlmeir, A.~Rueda, \emph{et~al.}, \enquote{Sensitivity limits of millimeter-wave photonic radiometers based on efficient electro-optic upconverters,} {\protect\JournalTitle{Optica}} \textbf{5}, 1210--1219 (2018).

\bibitem{abdalmalak_integrated_2022}
K.~A. Abdalmalak, G.~S. Botello, M.~I. Suresh, \emph{et~al.}, \enquote{An {Integrated} {Millimeter}-{Wave} {Satellite} {Radiometer} {Working} at {Room}-{Temperature} with {High} {Photon} {Conversion} {Efficiency},} {\protect\JournalTitle{Sensors}} \textbf{22}, 2400 (2022).

\bibitem{logan_triply-resonant_2023}
A.~D. Logan, S.~Shree, S.~Chakravarthi, \emph{et~al.}, \enquote{Triply-resonant sum frequency conversion with gallium phosphide ring resonators,} {\protect\JournalTitle{Optics Express}} \textbf{31}, 1516--1531 (2023).

\bibitem{suresh_multichannel_2025}
M.~I. Suresh, F.~Sedlmeir, D.~W. Vogt, \emph{et~al.}, \enquote{Multichannel upconversion of terahertz radiation in an optical disk resonator,} {\protect\JournalTitle{Optics Express}} \textbf{33}, 10302--10311 (2025).

\bibitem{schunk_identifying_2014}
G.~Schunk, J.~U. Fürst, M.~Förtsch, \emph{et~al.}, \enquote{Identifying modes of large whispering-gallery mode resonators from the spectrum and emission pattern,} {\protect\JournalTitle{Optics Express}} \textbf{22}, 30795--30806 (2014).

\bibitem{knight_mapping_1995}
J.~C. Knight, N.~Dubreuil, V.~Sandoghdar, \emph{et~al.}, \enquote{Mapping whispering-gallery modes in microspheres with a near-field probe,} {\protect\JournalTitle{Optics Letters}} \textbf{20}, 1515--1517 (1995).

\bibitem{schmidt_near-field_2012}
C.~Schmidt, M.~Liebsch, A.~Klein, \emph{et~al.}, \enquote{Near-field mapping of optical eigenstates in coupled disk microresonators,} {\protect\JournalTitle{Physical Review A}} \textbf{85}, 033827 (2012).

\bibitem{dev_near-field_2022}
S.~U. Dev, N.~M. Anthony, S.~Trendafilov, \emph{et~al.}, \enquote{Near-field mapping of high permittivity dielectric microwave resonator modes via optically induced conductance,} {\protect\JournalTitle{Optics Express}} \textbf{30}, 13583--13590 (2022).

\bibitem{hale_near-field_2023}
L.~L. Hale, T.~Siday, and O.~Mitrofanov, \enquote{Near-field imaging and spectroscopy of terahertz resonators and metasurfaces [{Invited}],} {\protect\JournalTitle{Optical Materials Express}} \textbf{13}, 3068--3086 (2023).

\bibitem{lee_terahertz_2017}
W.~S.~L. Lee, K.~Kaltenecker, S.~Nirantar, \emph{et~al.}, \enquote{Terahertz near-field imaging of dielectric resonators,} {\protect\JournalTitle{Optics Express}} \textbf{25}, 3756--3764 (2017).

\bibitem{mitrofanov_near-field_2018}
O.~Mitrofanov, Y.~Todorov, D.~Gacemi, \emph{et~al.}, \enquote{Near-field spectroscopy and tuning of sub-surface modes in plasmonic terahertz resonators,} {\protect\JournalTitle{Optics Express}} \textbf{26}, 7437--7450 (2018).

\bibitem{schiattarella_terahertz_2024}
C.~Schiattarella, A.~Di~Gaspare, L.~Viti, \emph{et~al.}, \enquote{Terahertz near-field microscopy of metallic circular split ring resonators with graphene in the gap,} {\protect\JournalTitle{Scientific Reports}} \textbf{14}, 16227 (2024).

\bibitem{foreman_dielectric_2016}
M.~R. Foreman, F.~Sedlmeir, H.~G.~L. Schwefel, and G.~Leuchs, \enquote{Dielectric tuning and coupling of whispering gallery modes using an anisotropic prism,} {\protect\JournalTitle{JOSA B}} \textbf{33}, 2177--2195 (2016).

\bibitem{azeem_dielectric_2021}
F.~Azeem, L.~S. Trainor, P.~A. Devane, \emph{et~al.}, \enquote{Dielectric perturbations: anomalous resonance frequency shifts in optical resonators,} {\protect\JournalTitle{Optics Letters}} \textbf{46}, 2477--2480 (2021).

\bibitem{vogt_anomalous_2019}
D.~W. Vogt, A.~H. Jones, H.~G.~L. Schwefel, and R.~Leonhardt, \enquote{Anomalous blue-shift of terahertz whispering-gallery modes via dielectric and metallic tuning,} {\protect\JournalTitle{Optics Letters}} \textbf{44}, 1319--1322 (2019).

\bibitem{suresh_gallium_2023}
M.~I. Suresh, H.~G.~L. Schwefel, and D.~W. Vogt, \enquote{Gallium arsenide whispering gallery mode resonators for terahertz photonics,} {\protect\JournalTitle{Optics Express}} \textbf{31}, 33056--33063 (2023).

\bibitem{ta_tuning_2013}
V.~D. Ta, R.~Chen, and H.~D. Sun, \enquote{Tuning {Whispering} {Gallery} {Mode} {Lasing} from {Self}-{Assembled} {Polymer} {Droplets},} {\protect\JournalTitle{Scientific Reports}} \textbf{3}, 1362 (2013). Publisher: Nature Publishing Group.

\bibitem{schunk_frequency_2016}
G.~Schunk, V.~, U., S.~, F., \emph{et~al.}, \enquote{Frequency tuning of single photons from a whispering-gallery mode resonator to {MHz}-wide transitions,} {\protect\JournalTitle{Journal of Modern Optics}} \textbf{63}, 2058--2073 (2016). Publisher: Taylor \& Francis \_eprint: https://doi.org/10.1080/09500340.2016.1148211.

\bibitem{munoz-hernandez_tunable_2019}
T.~Muñoz-Hernández, E.~Reyes-Vera, and P.~Torres, \enquote{Tunable {Whispering} {Gallery} {Mode} {Photonic} {Device} {Based} on {Microstructured} {Optical} {Fiber} with {Internal} {Electrodes},} {\protect\JournalTitle{Scientific Reports}} \textbf{9}, 12083 (2019). Publisher: Nature Publishing Group.

\bibitem{stanze_compact_2011}
D.~Stanze, A.~Deninger, A.~Roggenbuck, \emph{et~al.}, \enquote{Compact cw {Terahertz} {Spectrometer} {Pumped} at 1.5 $\mu$m {Wavelength},} {\protect\JournalTitle{Journal of Infrared, Millimeter, and Terahertz Waves}} \textbf{32}, 225--232 (2011).

\bibitem{deninger_275_2015}
A.~J. Deninger, A.~Roggenbuck, S.~Schindler, and S.~Preu, \enquote{2.75 {THz} tuning with a triple-{DFB} laser system at 1550 nm and {InGaAs} photomixers,} {\protect\JournalTitle{Journal of Infrared, Millimeter, and Terahertz Waves}} \textbf{36}, 269--277 (2015).

\bibitem{vogt_high_2017}
D.~W. Vogt and R.~Leonhardt, \enquote{High resolution terahertz spectroscopy of a whispering gallery mode bubble resonator using {Hilbert} analysis,} {\protect\JournalTitle{Optics Express}} \textbf{25}, 16860--16866 (2017).

\bibitem{mazzei_controlled_2007}
A.~Mazzei, S.~Götzinger, L.~de S. Menezes, \emph{et~al.}, \enquote{Controlled {Coupling} of {Counterpropagating} {Whispering}-{Gallery} {Modes} by a {Single} {Rayleigh} {Scatterer}: {A} {Classical} {Problem} in a {Quantum} {Optical} {Light},} {\protect\JournalTitle{Physical Review Letters}} \textbf{99}, 173603 (2007).

\bibitem{wang_direct_2021}
S.~Wang, S.~Liu, Y.~Liu, \emph{et~al.}, \enquote{Direct observation of chaotic resonances in optical microcavities,} {\protect\JournalTitle{Light: Science \& Applications}} \textbf{10}, 135 (2021).

\end{thebibliography}
\end{document}